\documentclass{article}
\usepackage{amssymb}
\usepackage{amsfonts}
\usepackage{amsmath}

\input{tcilatex}
\begin{document}

\title{Comment on " The Rotation-Vibration Spectrum of Diatomic Molecules
with the Tietz-Hua Rotating Oscillator "}
\author{A. Khodja, F. Benamira and L. Guechi \\
%EndAName
Laboratoire de Physique Th\'{e}orique, D\'{e}partement de Physique, \and %
Facult\'{e} des Sciences Exactes, Universit\'{e} des Fr\`{e}res Mentouri,
Constantine, \and Route d'Ain El Bey, Constantine, Algeria}
\maketitle

\begin{abstract}
We present arguments demonstrating that the application of the
Nikiforov-Uvarov polynomial method to solve the Schr\"{o}dinger equation
with the Tietz-Hua potential is valid only when $e^{-b_{h}r_{e}}\leq c_{h}<1$
and $r_{0}<r<+\infty $. In particular, it is point out that the numerical
results with $c_{h}\neq 0$\ for the diatomic molecules HF, N$_{2}$, I$_{2}$,
H$_{2}$, O$_{2}$ and O$_{2}^{+}$\ given in Tables 3-5 by Hamzavi and
co-workers are wrong. When $-1<c_{h}$ $<0$ or $0<c_{h}<e^{-b_{h}r_{e}},$
this approach is not suitable. In both cases, it is shown that the solutions
of the Schr\"{o}dinger equation are expressed in terms of the generalized
hypergeometric functions $_{2}F_{1}(a,b,c;z)$. The determination of the
energy levels requires the solution of transcendental equations involving
the hypergeometric function by means of the numerical procedure.

PACS: 03.65.-w, 03.65.Ge

Keywords: Schr\"{o}dinger equation; Nikiforov-Uvarov method; Tietz-Hua
potential; Morse potential; Bound states.
\end{abstract}

In a recent work \cite{Hamzavi} published in this journal, Hamzavi and
co-workers claimed to have obtained the vibrational energy levels
corresponding to $s$ states of a set of diatomic molecules through the
resolution of the radial Schr\"{o}dinger equation

\begin{equation}
\left[ \frac{d^{2}}{dr^{2}}+\frac{2\mu }{\hbar ^{2}}\left(
E-V_{TH}(r)\right) \right] R_{E,0}(r)=0,  \label{a.1}
\end{equation}%
where $\mu $ is the reduced mass of the rotating oscillator, $R_{E,0}(r)$
denotes the reduced radial\ wave function of a $s$ state and $V_{TH}(r)$ is
the so-called Tietz-Hua potential function \cite{Tietz,Hua,Natanzon} defined
by

\begin{equation}
V_{TH}(r)=D\left[ \frac{1-e^{-b_{h}(r-r_{e})}}{1-c_{h}e^{-b_{h}(r-r_{e})}}%
\right] ^{2};\text{ }b_{h}=\beta (1-c_{h}).  \label{a.2}
\end{equation}%
In equation (\ref{a.2}), $r$ is the internuclear distance, $D$ and $r_{e}$
stand for the potential well depth and the molecular bond length,
respectively. $\beta $ is the Morse constant and for physical reasons the
dimensionless constant $c_{h}$ is an optimization parameter chosen such that 
$\left\vert c_{h}\right\vert <1$.

To solve (\ref{a.1}), the authors of the Ref. \cite{Hamzavi} introduce the
new variable $s=e^{-\alpha x}$ with $\alpha =b_{h}r_{e}$ and $x=\frac{r-r_{e}%
}{r_{e}}$ and use the parametric generalization of polynomial
Nikiforov-Uvarov (NU) method \cite{Nikiforov} without considering of the
conditions of its application. One can see by simple inspection that the
Tietz-Hua potential (\ref{a.2}) is not continuous throughout the interval $%
%TCIMACRO{\U{211d} }%
%BeginExpansion
\mathbb{R}
%EndExpansion
^{+}$ whatever $c_{h}\in \left] -1,1\right[ $. It has a strong singularity
at the point $r=r_{0}=r_{e}+\frac{1}{b_{h}}\ln c_{h}$ when $c_{h}>0.$
Moreover, according to the theorem on the orthogonality of
hypergeometric-type polynomials (see Ref. \cite{Nikiforov}, Eq. (17), p.
29), we note that the weight function $\rho (s)$ satisfies the condition%
\begin{equation}
\left. \sigma (s)\rho (s)s^{k}\right\vert _{a}^{b}=0;\text{ \ \ }%
(k=0,1,2,...),  \label{a.3}
\end{equation}%
only in the case where $e^{-b_{h}r_{e}}\leq c_{h}<1$. Here $\left(
a,b\right) =\left( \frac{1}{c_{h}},0\right) $, and the polynomials $\sigma
(s)$ and $\rho (s)$ are given by (see Eqs. (A3) and (17) in Ref. \cite%
{Hamzavi})%
\begin{equation}
\sigma (s)=s(1-c_{h}s),  \label{a.4}
\end{equation}%
and%
\begin{equation}
\rho (s)=s^{2\sqrt{\frac{r_{e}^{2}}{\alpha ^{2}}(d-\varepsilon )}%
}(1-c_{h}s)^{2\left( \sqrt{\frac{1}{4}+\frac{r_{e}^{2}d}{c_{h}^{2}\alpha ^{2}%
}\left( c_{h}-1\right) ^{2}}-1\right) },  \label{a.5}
\end{equation}%
(note that $c_{h}$ is missing in the second factor of equation (17)). As the
expression (15) of the energy eigenvalues obtained in \cite{Hamzavi} is
incorrect, it is worthwhile to discuss again the resolution of the equation (%
\ref{a.1}) considering all the possible cases:

(i) $\ \ e^{-b_{h}r_{e}}\leq c_{h}<1$ and $r_{0}<r<+\infty $

By introducing the new variable

\begin{equation}
\widetilde{s}=c_{h}s,  \label{a.6}
\end{equation}
the radial Schr\"{o}dinger equation (\ref{a.1}) can be reduced to

\begin{equation}
\left[ \widetilde{s}(1-\widetilde{s})\frac{d^{2}}{d\widetilde{s}^{2}}+(1-%
\widetilde{s})\frac{d}{d\widetilde{s}}+\frac{1}{b_{h}^{2}}\left( \frac{%
\widetilde{d}}{c_{h}^{2}}-\varepsilon \right) -\frac{\widetilde{d}%
-\varepsilon }{b_{h}^{2}\widetilde{s}}-\frac{\widetilde{d}}{b_{h}^{2}}\frac{%
\left( 1-\frac{1}{c_{h}}\right) ^{2}}{1-\widetilde{s}}\right] R_{E,0}(%
\widetilde{s})=0,  \label{a.7}
\end{equation}%
where $\varepsilon =\frac{2\mu E}{\hbar ^{2}}$ and $\widetilde{d}=\frac{2\mu
D}{\hbar ^{2}}$. Since $\widetilde{s}=0$ and $\widetilde{s}=1$ are two
singularities of (\ref{a.7}), we look for a solution in the form

\begin{equation}
R_{E,0}(\widetilde{s})=\widetilde{s}^{\lambda }\left( 1-\widetilde{s}\right)
^{\delta }u_{E,0}(\widetilde{s}).  \label{a.8}
\end{equation}%
If we impose on $\lambda $ and $\nu $ the conditions%
\begin{equation}
\left\{ 
\begin{array}{c}
\lambda ^{2}=\frac{\widetilde{d}-\varepsilon }{b_{h}^{2}},\text{ } \\ 
\left( \delta -\frac{1}{2}\right) ^{2}=\frac{1}{4}+\frac{\widetilde{d}}{%
b_{h}^{2}}\left( 1-\frac{1}{c_{h}}\right) ^{2},%
\end{array}%
\right.  \label{a.9}
\end{equation}%
and on account of the boundary conditions

\begin{equation}
R_{E,0}(1)=0,  \label{a.10}
\end{equation}%
and%
\begin{equation}
R_{E,0}(0)=0,  \label{a.11}
\end{equation}%
both $\lambda $ and $\nu $ have to be positive. Substituting (\ref{a.8}) in (%
\ref{a.7}) and taking

\begin{equation}
\lambda =\frac{1}{b_{h}}\sqrt{\widetilde{d}-\varepsilon },  \label{a.12}
\end{equation}%
and%
\begin{equation}
\delta =\frac{1}{2}+\sqrt{\frac{1}{4}+\frac{\widetilde{d}}{b_{h}^{2}}\left(
1-\frac{1}{c_{h}}\right) ^{2}},  \label{a.13}
\end{equation}%
the following differential equation for $u_{E,0}(\widetilde{s})$ is obtained

\begin{equation}
\left\{ \widetilde{s}(1-\widetilde{s})\frac{d^{2}}{d\widetilde{s}^{2}}+\left[
2\lambda +1-\left( 2\lambda +2\delta +1\right) \widetilde{s}\right] \frac{d}{%
d\widetilde{s}}-\left( \lambda +\delta \right) ^{2}+\gamma ^{2}\right\}
u_{E,0}(\widetilde{s})=0,  \label{a.14}
\end{equation}%
with%
\begin{equation}
\gamma =\frac{1}{b_{h}}\sqrt{\frac{\widetilde{d}}{c_{h}^{2}}-\varepsilon }.
\label{a.15}
\end{equation}%
The solution of this equation, for which (\ref{a.8}) fulfills the boundary
condition (\ref{a.11}), can be written%
\begin{equation}
u_{E,0}(\widetilde{s})=N\text{ }_{2}F_{1}\left( \lambda +\delta -\gamma
,\lambda +\delta +\gamma ,2\lambda +1;\widetilde{s}\right) ,  \label{a.16}
\end{equation}%
where $N$ is a constant factor. Thus, the wave function satisfying Eqs. (\ref%
{a.1}), (\ref{a.8}) and (\ref{a.11}) is given by%
\begin{equation}
R_{E,0}(\widetilde{s})=N\widetilde{s}^{\lambda }(1-\widetilde{s})^{\delta }%
\text{ }_{2}F_{1}\left( \lambda +\delta -\gamma ,\lambda +\delta +\gamma
,2\lambda +1;\widetilde{s}\right) .  \label{a.17}
\end{equation}%
For the wave function to remain finite as $\widetilde{s}\rightarrow 0$, i.e. 
$r\rightarrow +\infty $, one has to have

\begin{equation}
\lambda +\delta -\gamma =-n_{r},  \label{a.18}
\end{equation}%
where $n_{r}=0,1,2,...,$ (the hypergeometric function reduces to a Jacobi
polynomial). The energy eigenvalues are then given by%
\begin{equation}
E_{n_{r},0}=D-\frac{\hbar ^{2}b_{h}^{2}}{8\mu }\left[ n_{r}+\delta -\frac{%
\frac{2\mu D}{\hbar ^{2}b_{h}^{2}}\left( \frac{1}{c_{h}^{2}}-1\right) }{%
n_{r}+\delta }\right] ^{2},  \label{a.19}
\end{equation}%
and the corresponding eigenfunctions by%
\begin{equation}
R_{n_{r},0}(r)=N_{n_{r}}\left[ c_{h}e^{-b_{h}(r-r_{e})}\right] ^{\lambda }%
\left[ 1-c_{h}e^{-b_{h}(r-r_{e})}\right] ^{\delta }P_{n_{r}}^{\left(
2\lambda ,2\delta -1\right) }\left( 1-2c_{h}e^{-b_{h}(r-r_{e})}\right) .
\label{a.20}
\end{equation}%
The $P_{n}^{(\alpha ,\beta )}$ are Jacobi polynomials. In Eq. (\ref{a.19}),
the normalization constant $N_{n_{r}}$ reads%
\begin{equation}
N_{n_{r}}=\left[ 2b_{h}\frac{\lambda (n_{r}+\lambda +\delta )}{n_{r}+\delta }%
\frac{n_{r}!\Gamma (n_{r}+2\lambda +2\delta )}{\Gamma (n_{r}+2\lambda
+1)\Gamma (n_{r}+2\delta )}\right] ^{\frac{1}{2}}.  \label{a.21}
\end{equation}%
The number of bound states $n_{r\max }$ is set by $n_{r\max }=\left\{ \frac{1%
}{\hbar b_{h}}\sqrt{2\mu D\left( \frac{1}{c_{h}^{2}}-1\right) }-\delta
\right\} ,$ and $\left\{ k\right\} $ denotes the largest integer inferior to 
$k$. Note that the numerical results of the energy levels for some molecules
can be calculated from expression (\ref{a.19}) when the values of the
parameter $c_{h}$ are greater than or equal to those contained in Table 1.

\begin{tabular}{|cccc|}
\hline
\multicolumn{4}{|c|}{Table 1: minimal values of the parameter $c_{h}$ for
obtaining the energy levels from Eq.(\ref{a.19}).} \\ \hline
molecule & \multicolumn{1}{|c}{\ \ \ $b_{h}\left( \mathring{A}^{-1}\right) $}
& \multicolumn{1}{|c}{\ \ $r_{e}\left( \mathring{A}\right) $} & 
\multicolumn{1}{|c|}{\ \ \ \ $c_{h}$} \\ \hline
HF & \ \ \ \ \ \ 1,94207\ \ \ \ \ \  & \ 0,917 & 0,168490115 \\ 
N$_{2}$\  & 2,78585 & 1,097 & 0,047071975 \\ 
I$_{2}$ & 2,12343 & 2,666 & 0,003478812 \\ 
H$_{2}$ \  & 1,61890 & 0,741 & 0,301313237 \\ 
O$_{2}$ & 2,59103 & 1,207 & 0,043832785 \\ 
O$_{2}^{+}$ \  & 2,86987 & 1,116 & 0,040649248 \\ \hline
\end{tabular}

(ii) $e^{-b_{h}r_{e}}\leqslant c_{h}<1$ and $0<r<r_{0}$

The solution of Eq. (\ref{a.1}) can not be obtained analytically and has no
physical interest.

(iii) $\ 0<c_{h}<e^{-b_{h}r_{e}}$ and $r\in 
%TCIMACRO{\U{211d} }%
%BeginExpansion
\mathbb{R}
%EndExpansion
^{+}$

The analysis presented above holds. But in this case, by using the boundary
condition $R_{E,0}(r)\underset{r\rightarrow +\infty }{\rightarrow }0$, we
show that the solution of the radial Schr\"{o}dinger equation (\ref{a.1})
can be written as

\begin{equation}
R_{E,0}(r)=\left[ c_{h}e^{-b_{h}(r-r_{e})}\right] ^{\lambda }\left[ \varphi
^{+}(r)+\varphi ^{-}(r)\right] ,  \label{a.22}
\end{equation}%
with%
\begin{eqnarray}
\varphi ^{\pm }(r) &=&C^{\pm }\left[ 1-c_{h}e^{-b_{h}(r-r_{e})}\right]
^{\delta _{\pm }}\text{ }  \notag \\
&&\times \text{ }_{2}F_{1}\left( \lambda +\delta _{\pm }-\gamma ,\lambda
+\delta _{\pm }+\gamma ,2\lambda +1;c_{h}e^{-b_{h}(r-r_{e})}\right) ,
\label{a.23}
\end{eqnarray}%
where $C^{\pm }$ are two constant factors and 
\begin{equation}
\delta _{\pm }=\frac{1}{2}\pm \sqrt{\frac{1}{4}+\frac{\widetilde{d}}{%
b_{h}^{2}}\left( 1-\frac{1}{c_{h}}\right) ^{2}}.  \label{a.24}
\end{equation}%
Now, taking into account the formula (see Ref. \cite{Gradshtein}, Eq.
(9.131), p. 1043)%
\begin{equation}
_{2}F_{1}\left( a,b,c;z\right) =\left( 1-z\right) ^{c-a-b}\text{ }%
_{2}F_{1}\left( c-a,c-b,c;z\right) ,  \label{a.25}
\end{equation}%
and since $\delta _{-}=1-\delta _{+}$, we can rewrite the obtained bound
state wave functions (\ref{a.22}) as 
\begin{eqnarray}
R_{E,0}(r) &=&C\left[ c_{h}e^{-b_{h}(r-r_{e})}\right] ^{\lambda }\left[
1-c_{h}e^{-b_{h}(r-r_{e})}\right] ^{\delta _{+}}\text{ }  \notag \\
&&\times \text{ }_{2}F_{1}\left( \lambda +\delta _{+}-\gamma ,\lambda
+\delta _{+}+\gamma ,2\lambda +1;c_{h}e^{-b_{h}(r-r_{e})}\right) ,
\label{a.26} \\
&&  \notag
\end{eqnarray}%
where $C$ is a constant factor. This solution fulfills the boundary
condition $R_{E,0}(0)=0$, when%
\begin{equation}
_{2}F_{1}\left( \lambda +\delta _{+}-\gamma ,\lambda +\delta _{+}+\gamma
,2\lambda +1;c_{h}e^{b_{h}r_{e}}\right) =0\text{.}  \label{a.27}
\end{equation}%
It follows that the energy eigenvalues of the bound states can be found by
numerically solving the transcendental equation (\ref{a.27}).

(iv) $-1<c_{h}<0$ and $r\in 
%TCIMACRO{\U{211d} }%
%BeginExpansion
\mathbb{R}
%EndExpansion
^{+}$

In this case too, instead of $r$, let us introduce a new variable $s$
defined by

\begin{equation}
s=\frac{\left\vert c_{h}\right\vert e^{b_{h}r_{e}}}{e^{b_{h}r}+\left\vert
c_{h}\right\vert e^{b_{h}r_{e}}}.  \label{a.28}
\end{equation}%
By making the substitution%
\begin{equation}
R_{E,0}(r)=s^{\lambda }\left( 1-s\right) ^{\gamma }u_{E,0}(s),  \label{a.29}
\end{equation}%
and, with arguments similar to those used in the preceding case, we show
that the solution of Eq. (\ref{a.1}) has the form

\begin{eqnarray}
R_{E,0}(r) &=&C\left[ \frac{\left\vert c_{h}\right\vert }{%
e^{b_{h}(r-r_{e})}+\left\vert c_{h}\right\vert }\right] ^{\lambda }\left[ 
\frac{1}{1+\left\vert c_{h}\right\vert e^{-b_{h}(r-r_{e})}}\right] ^{%
\overline{\gamma }_{+}}\text{ }  \notag \\
&&\times \text{ }_{2}F_{1}\left( 1+\lambda +\overline{\gamma }_{+}-\overline{%
\delta }_{+},\lambda +\overline{\gamma }_{+}+\overline{\delta }_{+},2\lambda
+1;\frac{\left\vert c_{h}\right\vert }{e^{b_{h}(r-r_{e})}+\left\vert
c_{h}\right\vert }\right) ,  \notag \\
&&  \label{a.30}
\end{eqnarray}%
where%
\begin{equation}
\left\{ 
\begin{array}{c}
\overline{\delta }_{+}=\frac{1}{2}+\sqrt{\frac{1}{4}+\frac{\widetilde{d}}{%
b_{h}^{2}}\left( 1+\frac{1}{\left\vert c_{h}\right\vert }\right) ^{2}}, \\ 
\overline{\gamma }_{+}=\frac{1}{b_{h}}\sqrt{\frac{\widetilde{d}}{c_{h}^{2}}%
-\varepsilon },%
\end{array}%
\right.  \label{a.31}
\end{equation}%
and $C$ is a constant factor. As we see, the wave functions (\ref{a.30})
satisfy the boundary condition $R_{E,0}(0)=0,$ when%
\begin{equation}
_{2}F_{1}\left( 1+\lambda +\overline{\gamma }_{+}-\overline{\delta }%
_{+},\lambda +\overline{\gamma }_{+}+\overline{\delta }_{+},2\lambda +1;%
\frac{\left\vert c_{h}\right\vert }{e^{-b_{h}r_{e}}+\left\vert
c_{h}\right\vert }\right) =0.  \label{a.32}
\end{equation}%
Thus, the levels of energy bound states are determined by the solutions of
the transcendental equation (\ref{a.32}), which can be solved numerically.

(v) $c_{h}\rightarrow 0$ and $r\in 
%TCIMACRO{\U{211d} }%
%BeginExpansion
\mathbb{R}
%EndExpansion
^{+}$

If we let $c_{h}\rightarrow 0$, the expression (\ref{a.2}) reduces to the
radial Morse potential%
\begin{equation}
V_{M}(r)=D\left[ 1-e^{-\beta (r-r_{e})}\right] ^{2}.  \label{a.33}
\end{equation}

In this case, we can see from (\ref{a.12}), (\ref{a.15}) and (\ref{a.24})
that

\begin{equation}
\left\{ 
\begin{array}{c}
\lambda \underset{c_{h}\rightarrow 0}{\simeq }\frac{1}{\hbar \beta }\sqrt{%
2\mu \left( D-E\right) }, \\ 
\gamma \underset{c_{h}\rightarrow 0}{\simeq }\frac{\sqrt{2\mu D}}{\hbar
\beta c_{h}}\rightarrow +\infty \\ 
\delta _{+}\underset{c_{h}\rightarrow 0}{\simeq }\frac{1}{2}+\frac{\sqrt{%
2\mu D}}{\hbar \beta }\left( \frac{1}{c_{h}}-1\right) \rightarrow +\infty .%
\end{array}%
\right.  \label{a.34}
\end{equation}

On the other hand, using the relation of the confluent hypergeometric
function to the hypergeometric series \cite{Landau}%
\begin{equation}
\text{ }_{1}F_{1}\left( \alpha ,\gamma ;z\right) =\underset{\beta
\rightarrow \infty }{\lim }\text{ }_{2}F_{1}\left( \alpha ,\beta ,\gamma ;%
\frac{z}{\beta }\right) ,  \label{a.35}
\end{equation}%
we can show without difficulty that, as $c_{h}\rightarrow 0,$ the wave
functions (\ref{a.26}) can be expressed as:%
\begin{eqnarray}
&&R_{E,0}(r)\underset{c_{h}\rightarrow 0}{\simeq }Ne^{-\frac{\sqrt{2\mu
\left( D-E\right) }}{\hbar }\left( r-r_{e}\right) }\exp \left[ -\frac{\sqrt{%
2\mu D}}{\hbar \beta }e^{-\beta \left( r-r_{e}\right) }\right]  \notag \\
&&\times \text{ }_{1}F_{1}\left( \frac{1}{2}+\frac{1}{\beta \hbar }\left[ 
\sqrt{2\mu \left( D-E\right) }-\sqrt{2\mu D}\right] ,\frac{2\sqrt{2\mu
\left( D-E\right) }}{\hbar \beta }+1;\frac{2\sqrt{2\mu D}}{\hbar \beta }%
e^{-\beta \left( r-r_{e}\right) }\right) ,  \notag \\
&&  \label{a.36}
\end{eqnarray}%
where $N$ is a normalization constant. This solution vanishes at infinity
only if

\begin{equation}
\frac{1}{2}+\frac{1}{\beta \hbar }\left[ \sqrt{2\mu \left( D-E\right) }-%
\sqrt{2\mu D}\right] =-n_{r}\text{.}  \label{a.37}
\end{equation}%
Finally, from this condition we find the energy levels to be given by

\begin{equation}
E_{n_{r}}=D-\frac{\hbar ^{2}\beta ^{2}}{2\mu }\left( n_{r}+\frac{1}{2}-\frac{%
\sqrt{2\mu D}}{\hbar \beta }\right) ^{2},\text{ \ }n_{r}=0,1,2,...,\left\{ 
\frac{\sqrt{2\mu D}}{\hbar \beta }-\frac{1}{2}\right\} \text{.}  \label{a.38}
\end{equation}

Note that, by starting from the wave functions (\ref{a.30}) and proceeding
to the limit $\left\vert c_{h}\right\vert \rightarrow 0$, we recover the
wave functions (\ref{a.36}) and the energy spectrum (\ref{a.38}) of diatomic
molecules in the radial Morse potential.

In conclusion, the analytical and numerical results obtained by the authors
of Ref. \cite{Hamzavi} are inconsistent because \ the NU polynomial method
is used without taking into account the conditions for its application. The
radial Schr\"{o}dinger equation (\ref{a.1}) can only solved by this method
when $e^{-b_{h}r_{e}}\leqslant c_{h}<1$ and $r_{0}<r<+\infty $ or $c_{h}=0$
and $r\in 
%TCIMACRO{\U{211d} }%
%BeginExpansion
\mathbb{R}
%EndExpansion
^{+}$. It is quite evident that the potential (\ref{a.1}) \ in these two
cases corresponds to eigenfunctions which are represented by Jacobi
polynomials and confluent hypergeometric functions or Laguerre polynomials
respectively. Unfortunately, for $-1<c_{h}<0$ or $0<c_{h}<e^{-b_{h}r_{e}}$,
the NU method cannot be applied. This is a conceptual drawback of this
technique to treat wave equations admitting only orthogonal polynomials as
solutions. In the latter two cases, the exact solutions of Eq. (\ref{a.1})
are expressed in terms of hypergeometric series. From these, we have shown
by applying the boundary conditions that the energy levels can be found from
numerical solution of transcendental equations involving the hypergeometric
function.

\end{document}